\documentclass{elsart}

\usepackage{epsfig}
\usepackage{amssymb}

\def \beq {\begin{equation}}
\def \eeq {\end{equation}}

\begin{document}

\begin{frontmatter}

\title{On the viability of local criteria for chaos}

\author{Alberto Saa}
\ead{asaa@ime.unicamp.br}
\address{Departamento de Matem\'atica Aplicada, \\
IMECC -- UNICAMP, C.P. 6065, 13083-970 Campinas, SP, Brazil.}

\begin{abstract}
We consider here a recently proposed 
geometrical criterion for local
instability based on the geodesic deviation equation.
Although such a criterion can be useful in some cases,
we show here that, in general, it is 
neither necessary nor sufficient for the
occurrence of chaos. To this purpose, we introduce a
class of chaotic two-dimensional systems with Gaussian
curvature everywhere positive and, hence,
locally stable. We show explicitly that chaotic behavior
arises from some trajectories that
reach certain non convex parts of 
the boundary of the effective Riemannian manifold.
Our result questions, once more, the viability of
local, curvature-based criteria to predict chaotic behavior.

\end{abstract}

\end{frontmatter}

\section{Introduction}

In recent years, many efforts have been devoted  
to formulate a local geometrical
criterion which could indicate chaotic behavior in a mechanical
system. The main idea behind such attempts is to
provide the Lagrangian or Hamiltonian formulation of the equations
of motion with an
effective Riemaniann structure, and then search for some criterion
based on its curvature properties which could   predict
local instabilities that might lead to  
chaos\cite{krylov,arnold}. This line of research
has been mainly inspired by the celebrated result due to
Anosov\cite{avez} stating that the geodesic flow  is chaotic 
in compact manifolds for which all
sectional curvatures are everywhere negative.

In a recent work, Zscz\c esny and Dobrowolski\cite{ZD}
rederive in a straightforward way the relevant results necessary
to formulate such criteria.
We adopt here their conventions, which 
we will briefly recall now. Let us consider a classical mechanical
system with $N$ degrees of freedom described by the Lagrangian
\beq
\label{lagr}
{\mathcal L}(q,\dot{q}) = \frac{1}{2} g_{ij}(q)\dot{q}^i \dot{q}^j -
V(q),
\eeq 
where $i,j=1, 2,\dots, N$, 
the dot stands
to differentiation with respect to the time $t$, and
$g_{ij}$ is the Riemannian
metric on the configuration space $\mathcal M$. All the
quantities here are assumed to be sufficiently 
smooth. The Euler-Lagrange
equations of (\ref{lagr}) can be written as
\beq
\label{EL}
\ddot{q}^i + \Gamma^i_{jk} \dot{q}^j \dot{q}^k = - 
g^{ij}\partial_j V(q),
\eeq
where $\Gamma^i_{jk}$ is the Levi-Civita connection for the
metric $g_{ij}$.

The Hamiltonian of the system described by (\ref{lagr})
\beq
\label{ham}
{\mathcal H}(q,p) = \frac{1}{2} g^{ij}(q){p}_i {p}_j +
V(q),
\eeq
with $p_i=g_{ij}\dot{q}^j$,
is obviously a constant of motion, namely the total energy.
For a fixed energy $E$, the trajectories in the phase-space
are confined to the hypersurface
$E = \frac{1}{2} g^{ij}{p}_i {p}_j + V(q).$
The admissible region for the trajectories in the configuration
space is, therefore, given by
\beq
{\mathcal D}_E = \{q\in{\mathcal M} : V(q) \le E \}.
\eeq
In general, the region ${\mathcal D}_E$ can be bounded or not, connected
or not. The
boundary of the admissible region for the trajectories
is given by
\beq
\partial {\mathcal D}_E = \{q\in{\mathcal M} : V(q) = E \}.
\eeq
If the potential has no critical points on the boundary
$(\nabla V\ne 0)$, then
$\partial {\mathcal D}_E $ is a ($N-1$)-dimensional submanifold
of $\mathcal M$.
We can easily see that if a trajectory reaches the boundary
$\partial {\mathcal D}_E $ at a point $q_0$, its velocity at
this point vanishes and the trajectory approach
or depart from $q_0$ perpendicularly
to the boundary $\partial {\mathcal D}_E $. In particular,
there is no allowed trajectory along the boundary.

The crucial observation here is that the equations of
motion (\ref{EL}) are, in the interior of
${\mathcal D}_E$,  fully equivalent to the geodesic equation
in the ``effective'' 
Riemannian geometry on $\mathcal M$ defined by the Jacobi
metric\cite{ZD} 
\beq
\label{jacobi}
\hat{g}_{ij}(q) = 2(E-V(q)) g_{ij}(q).
\eeq
The geodesic equation for the effective geometry defined
from (\ref{jacobi}) is given by
\beq
\label{geo}
\hat{\nabla}_u u =
\frac{d^2 q^i}{ds^2} + \hat{\Gamma}^i_{jk}
\frac{dq^j}{ds} \frac{dq^k}{ds} = 0,
\eeq
where 
$u=dq^i/ds$ is the tangent vector along the geodesic, 
$\hat{\nabla}$ and 
$ \hat{\Gamma}^i_{jk}$ are, respectively, the covariant derivative
and the Levi-Civita connection
for the Jacobi metric $\hat{g}_{ij}$, and $s$ is a parameter
along the geodesic obeying
\beq
\label{para}
\frac{ds}{dt} = 2 (E-V(q)).
\eeq
It can be shown that $s(t)$ is a monotonically increasing 
function of time in the interior of ${\mathcal D}_E$\cite{ZD}. However,
from (\ref{para}) we can see that
$s$ is not a good evolution parameter on the boundary,
since $ds/dt=0$ on $\partial {\mathcal D}_E$.
As we will see, it is namely this 
failure, on the boundary $\partial {\mathcal D}_E$, 
 of the equivalence between the
equation of motion (\ref{EL}) and the geodesic equation
in the effective Riemannian manifold (\ref{geo}) that precludes
the possibility of obtaining an efficacious local, curvature-based
criterion to predict chaotic behavior of the system
governed by (\ref{EL}).

In a Riemannian geometry, the local tendency of geodesics
to converge or to diverge is governed by the geodesic
deviation equation:
\beq
\label{gde}
\hat{\nabla}_u \hat{\nabla}_u n = \hat{R}(u,n)u, 
\eeq
where $\hat{R}(u,n)$ is the 
Riemannian curvature of the Jacobi metric (\ref{jacobi}).
The vector field $n$, called
geodesic deviation, is orthogonal to $u$ and  
measures the distance between
nearby geodesics. For two-dimensional systems, one
can choose Fermi frames $(E_1,E_2)$ such that 
$u=E_1$ and $n= xE_2$, and the geodesic deviation equation
takes the simple form\cite{ZD}
\beq
\label{gde1}
\frac{d^2x}{ds^2} = - \hat{K}x,
\eeq 
where $\hat{K}$ is the Gaussian curvature of the two-dimensional 
Jacobi metric (\ref{jacobi}), which is given
by
\beq
\hat{K} = K + \frac{1}{4}\frac{\nabla^2 V}{(E-V)^2} + 
\frac{1}{4}\frac{(\nabla V)^2}{(E-V)^3},
\eeq
where $K$ is the Gaussian curvature of the 
original two-dimensional
configuration space $\mathcal M$. The Laplacian
$\nabla^2 V$ and the quadratic term 
$(\nabla V)^2$ are defined with respect to
the configuration space metric $g_{ij}$. Note that, with
the hypothesis that the potential is smooth 
($\nabla^2 V$ is finite) and has no critical points 
($\nabla V\ne 0$) on
$\partial {\mathcal D}_E$, the 
effective Gaussian curvature  $\hat{K}$ 
diverges on the boundary. Moreover, since $E-V>0$ inside
$ {\mathcal D}_E$,  $\hat{K}$ always assume arbitrary large positive
values near the boundary.
It is clear from (\ref{gde1}) that when the Gaussian
curvature is positive the geodesics converge locally.
On the other hand, if it is negative,
one has the local divergence of nearby geodesics. 
Zscz\c esny and  Dobrowolski illustrated the case of positive 
Gaussian curvature $\hat{K}$
with three known integrable 2-dimensional systems: quadratic potentials, 
the Toda lattice and the bounded Kepler problem.

If the admissible region for the trajectories on the configuration
space has no boundary, the equivalence between (\ref{EL}) and
(\ref{geo}) is complete. If, besides, all sectional curvatures
are everywhere negative, close trajectories governed by 
(\ref{EL}) tend to diverge at every point. However, even in such
case, it does not mean that the system is chaotic, as
Zscz\c esny and Dobrowolski have shown with their unbounded Kepler 
problem example. 
The key requirement of compactness is lacking here to reproduce
the result of Anosov. It is quite easy, indeed, to
generate integrable models with effective Gaussian curvature
everywhere negative on a non-compact $ {\mathcal D}_E$
by exploring harmonic potentials in higher dimensions\cite{ZD}.
The local divergence of nearby
trajectories is not, therefore, a sufficient condition to
the occurrence of chaos. 

In this work, we question again the viability of such local criteria
by showing, by means of an explicit example, that they are neither
necessary to the occurrence of chaos. In the next section,
we will introduce a two-dimensional system with the Gaussian
curvature $\hat{K}$
everywhere positive 
in the interior
of ${\mathcal D}_E$, and, hence, locally stable.
As we will see, chaotic behavior is indeed ruled out for large
sets of  trajectories 
reaching the boundary $\partial {\mathcal D}_E$. 
However, we identify a set of trajectories 
that reach a non convex part of 
the boundary and give rise to chaotic motion.
The situation here is similar to case of classical billiards\cite{billiard}.
Segments of trajectories between two successive bounces on the
boundary $\partial {\mathcal D}_E$ are integrable, but the
smooth match of an infinite number of some of these segments give rises
to a chaotic trajectory. As, in general, we cannot 
decide a priori if a given trajectory does or does not 
reach
the non convex part of 
boundary for some $t\in(-\infty, \infty)$, the positivity
of $\hat{K}$ inside 
${\mathcal D}_E$ is useless to predict if such trajectory is regular or not.

We notice that
the potential problems with the interpretation of the true motion
governed by (\ref{EL})
as a geodesic equation in an effective Riemannian manifold
with boundary were already pointed out in \cite{BT,SS,S}.
Some suggestions to avoid them based on distinct differential
spaces instead of differential manifolds were proposed in
\cite{SHS}.
We also notice that the power of local criteria to predict
chaos in General Relativity has also been questioned. 
Vieira and Letelier\cite{VL}, for instance, 
showed that the
local criterion based on the eigenvalues of the
Weyl curvature tensor presented by
Sota, Suzuki and Maeda\cite{SSM} were neither necessary nor
sufficient for the prediction of chaos in vacuum static axisymmetric
spacetimes.

Despite the fact that such curvature-based criteria are neither
necessary nor sufficient to the occurrence of chaos, the
analysis of the geodesic deviation equation for the Jacobi
metric has proved to be valuable in some cases. For instance,
in \cite{E-Z}, some chaotic behavior, confirmed by the numerical
evaluation of Lyapunov exponents, is identified in a Newtonian
gravitational problem with negative effective 
Gaussian curvature. In \cite{BS}, chaotic mixing behavior in
thermal-equilibrium beams, confirmed
also by some numerical calculations, is again identified in regions with
negative Gaussian curvature. In the other hand, regions of
regular motion in the anisotropic Manev problem\cite{DS} and
the Yang-Mills-Higgs\cite{DS1} system coincide with positive Gaussian 
curvature regions. We emphasize, however, that since the criteria based 
on the geodesic deviation 
equation are neither necessary nor sufficient for the occurrence of 
chaos, they must be used merely as {\em preliminary indicators} of 
integrability or chaos in dynamical systems.

\section{The model}

Our model is quite simple. It consists in a  
point-like particle moving under the potential corresponding
the superposition
of a quadratic attractive potential and some two-dimensional harmonic
repulsive monopoles. The configuration space $\mathcal M$ is
locally just $R^2$ with the Euclidean metric.
Let us consider explicitly
the simplest case with only one repulsive monopole
with charge $q$, placed at a distance $a$ from the center of the
quadratic potential.
The particle motion is governed by the Hamiltonian
${\mathcal H} = \frac{1}{2m} p_x^2 +
\frac{1}{2m}p_y^2 + V(x,y)$, with
\beq
\label{VV}
V(x,y) =  k\frac{(x-a)^2}{2} +  k\frac{{y}^2}{2} - 
\frac{q}{2}\ln \left( \frac{x^2 + y^2}{r_0^2}\right) 
\eeq
The parameter $r_0$ has no dynamical consequences, it
is chosen in order to guarantee that
the minimum of $V(x,y)$ be zero.
Clearly, in the limit $a\rightarrow 0$ we have a central-force
integrable model. 
The equations of motion are very simple. By introducing some
dimensionless quantities ($x\rightarrow ax$, $y\rightarrow ay$, 
$t\rightarrow \sqrt{m/k}t$), one has
\begin{eqnarray}
\label{ham1}
&p_x = \dot{x}&, \quad \dot{p}_x = -x\left(1-\frac{\alpha}{x^2 + y^2} \right) 
+ 1, 
\nonumber \\
&p_y = \dot{y}&, \quad \dot{p}_y = -y\left(1-\frac{\alpha}{x^2 + y^2} \right), 
\end{eqnarray}
where $\alpha = q/ka^2$.

\begin{figure}[pht]
\epsfxsize=7.0cm
\epsfbox{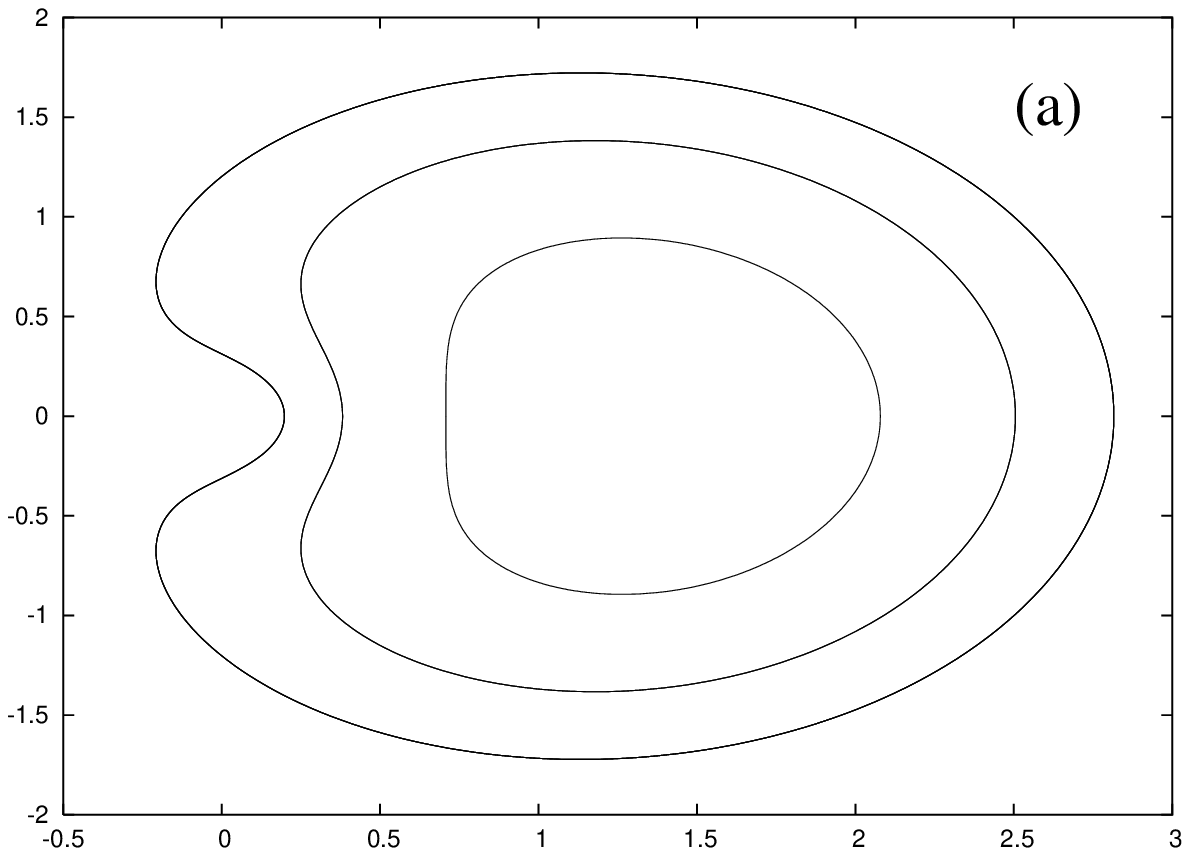}\hfill
\epsfxsize=7.0cm
\epsfbox{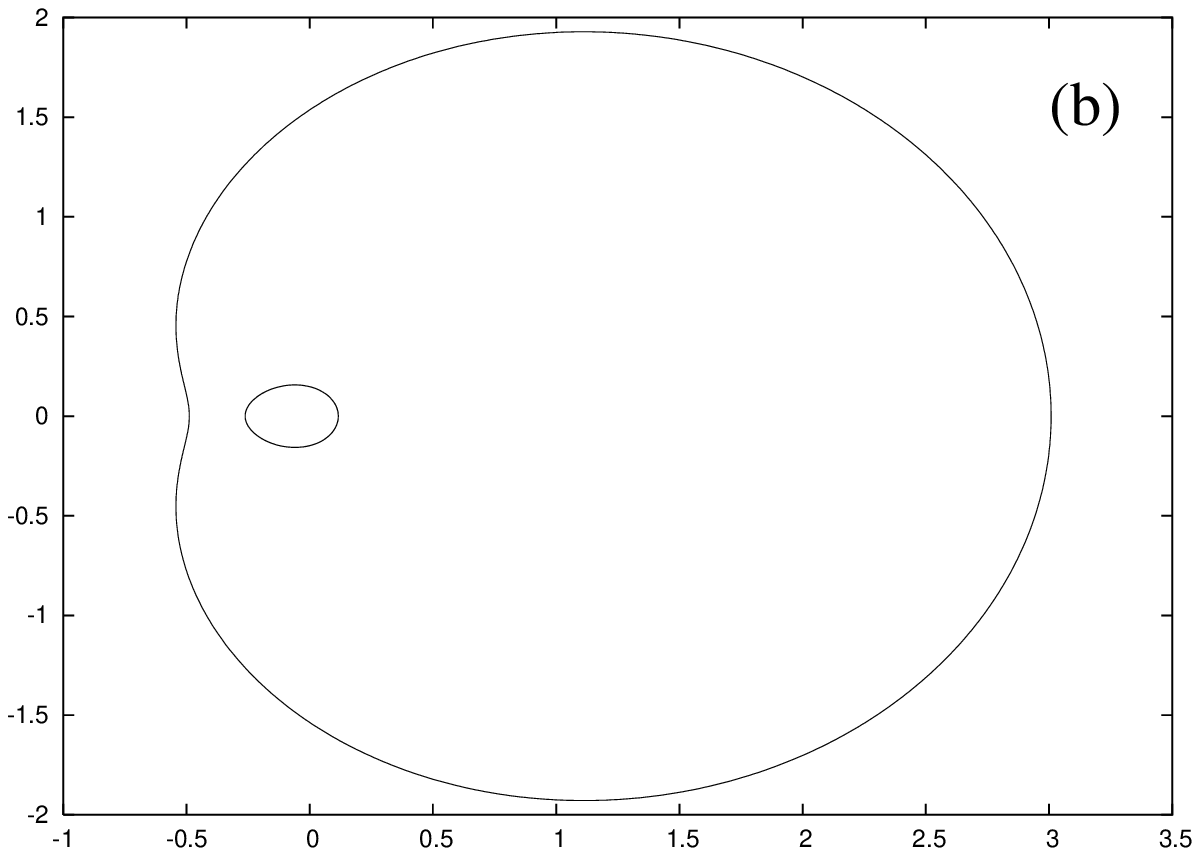}
\caption{The admissible region ${\mathcal D}_E \in R^2$ for the trajectories:
(a) the simply-connected case $E<E_{\rm C}$ and (b) the
multiply-connected one $E>E_{\rm C}$. The boundaries shown in 
(a) correspond  to the cases $V/E_{\rm C}=1/5, 1/2$ and $4/5$, while
in (b) $V/E_{\rm C}=1.02$. For all cases, $\alpha=1/2$.
The non convex parts of the boundaries, including the interior part for the
multiply-connected case, are the responsible for the bouncing of chaotic
trajectories.}
\end{figure}
We have two kinds of admissible
regions for the trajectories. For $E<E_{\rm C}$, where
\beq
E_{\rm C} = \frac{\beta}{2} + 
\alpha\ln \frac{\beta+1}{\beta-1}, 
\eeq
with $\beta=\sqrt{1+4\alpha}$,
${\mathcal D}_E$ is simply-connected, in contrast with the case for 
$E>E_{\rm C}$ (See Fig. 1). In both cases, $V(x,y)$ is smooth
in ${\mathcal D}_E$ and has no critical points on
$\partial {\mathcal D}_E$.

In the interior of ${\mathcal D}_E$, 
the Gaussian curvature $\hat{K}$ for this
system,
\beq
\hat{K} = \frac{1}{2(E-V(x,y))^2} + 
\frac{\left( x\left( 1 -\frac{\alpha}{x^2 + y^2} \right) -1 \right)^2
+ y^2 \left( 1 -\frac{\alpha}{x^2 + y^2} \right)^2 }{4(E-V(x,y))^3},
\eeq
is everywhere positive. 
Therefore, 
the nearby trajectories of our model approach each other at all points of
the interior of ${\mathcal D}_E$. As expected, the Gaussian curvature
diverges on the boundary.

\section{The dynamics}

In order to study the phase space of our system, we 
solve numerically the system governed by (\ref{ham1}) and construct
Poincar\'e's sections by using Henon's trick\cite{henon}.
We could do it very accurately, with a cumulative error,
measured by the constant Hamiltonian $\mathcal H$, inferior to $10^{-12}$.
Our results are shown in the next figures.

\begin{figure}[hpt]
\epsfxsize=10cm
\epsfbox{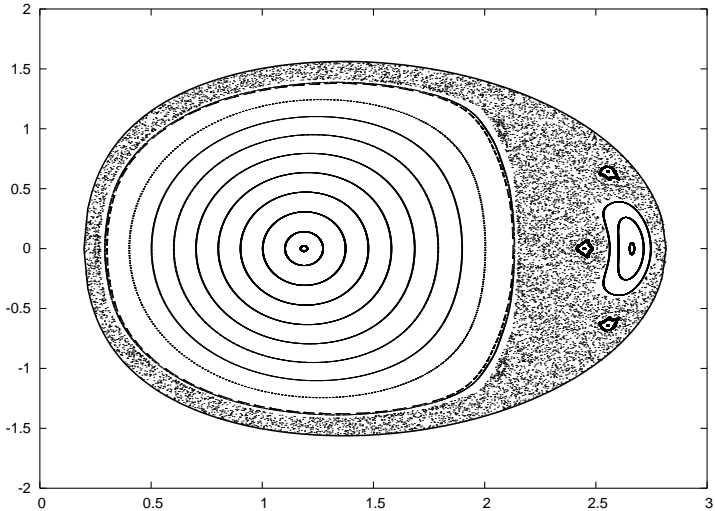}
\caption{Typical Poincar\`e's section $(x,p_x)$ across the
plane $y=0$ for the simply-connected ${\mathcal D}_E$ case.
For this section, one has $\alpha=1/2$ and $E/E_{\rm C}=4/5$.}
\end{figure}

In Fig. (2), we have a typical  Poincare's section across the plane
$y=0$ for the the simply-connected ${\mathcal D}_E$ case.
The regular 
solutions correspond to trajectories that do not probe the non convex
part of the boundary, see Fig. (3). Despite our exhaustive simulations,
we could not find a single trajectory that never touches the
boundary $\partial{\mathcal D}_E$. We were not able to prove
analytically the absence of such trajectories, but we notice that
the argument used in \cite{DS} to prove the existence of them does
not hold here.
\begin{figure}[ht]
\epsfxsize=7cm
\epsfbox{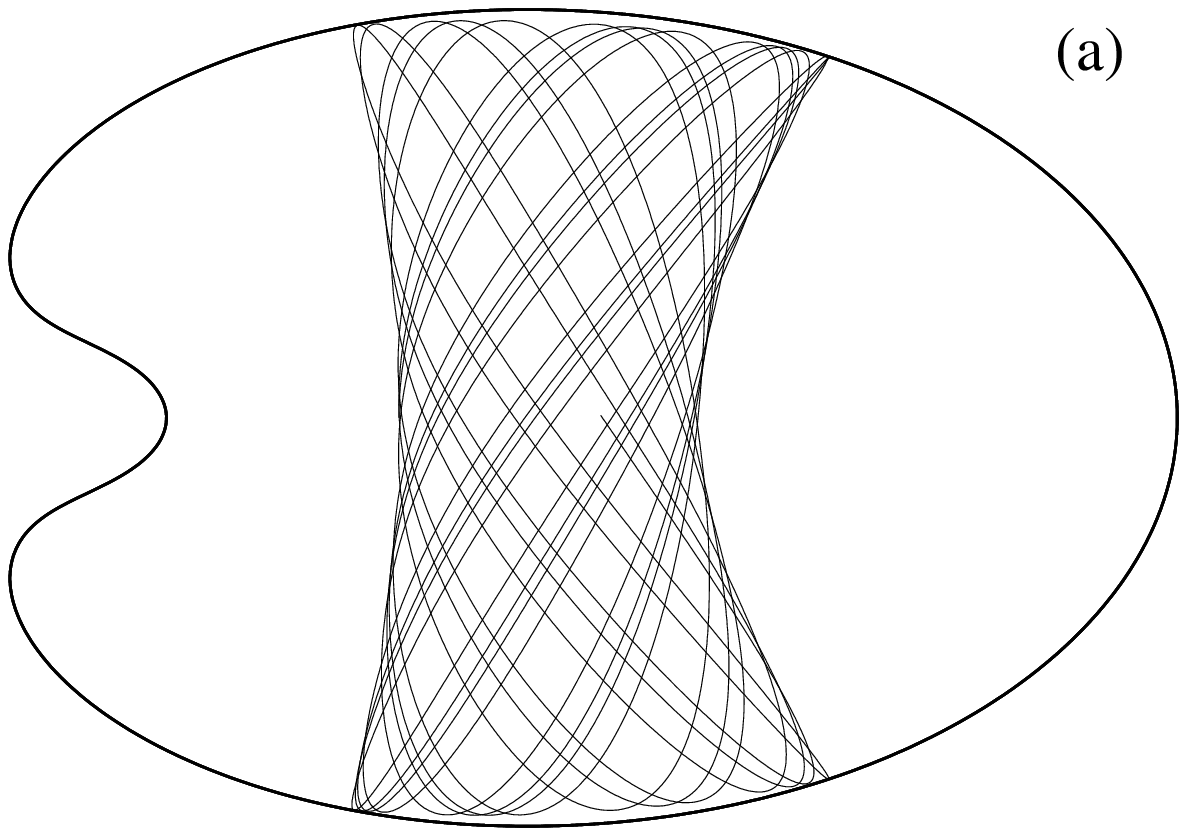} \hfill
\epsfxsize=7cm
\epsfbox{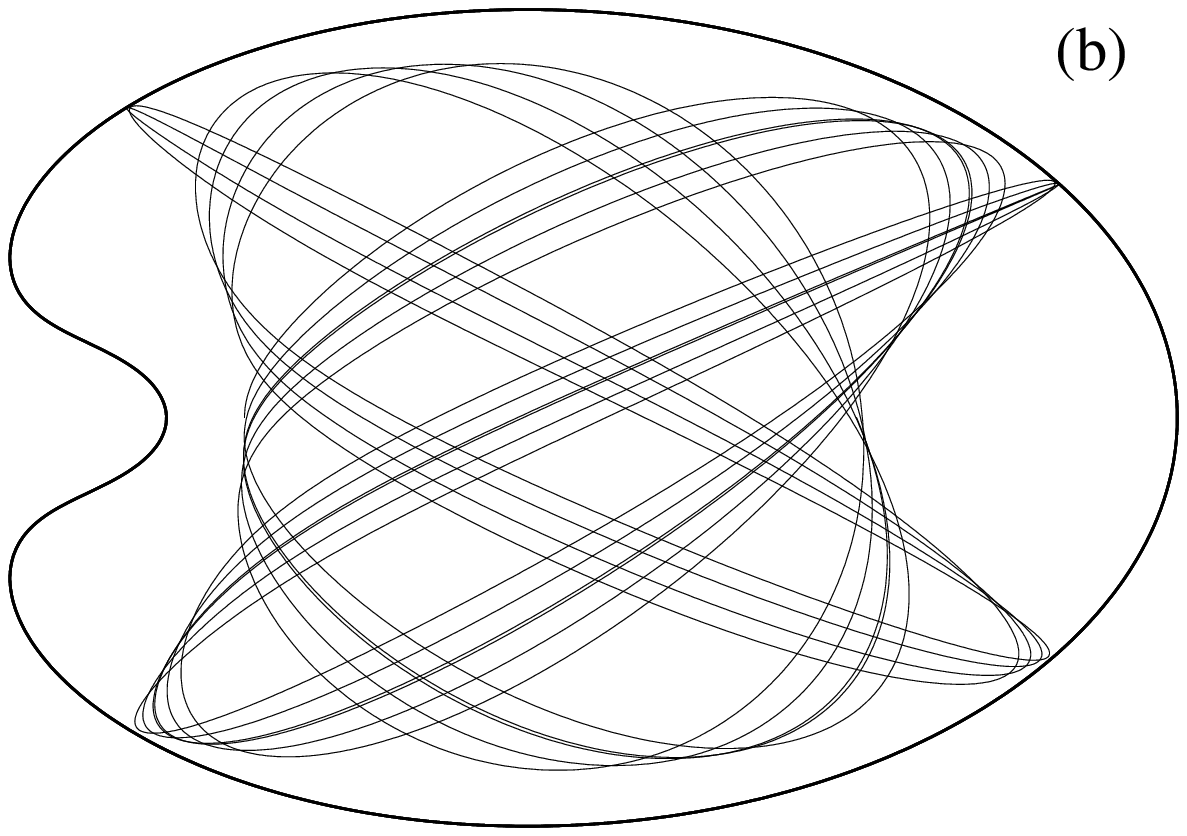} \\
\epsfxsize=7cm
\epsfbox{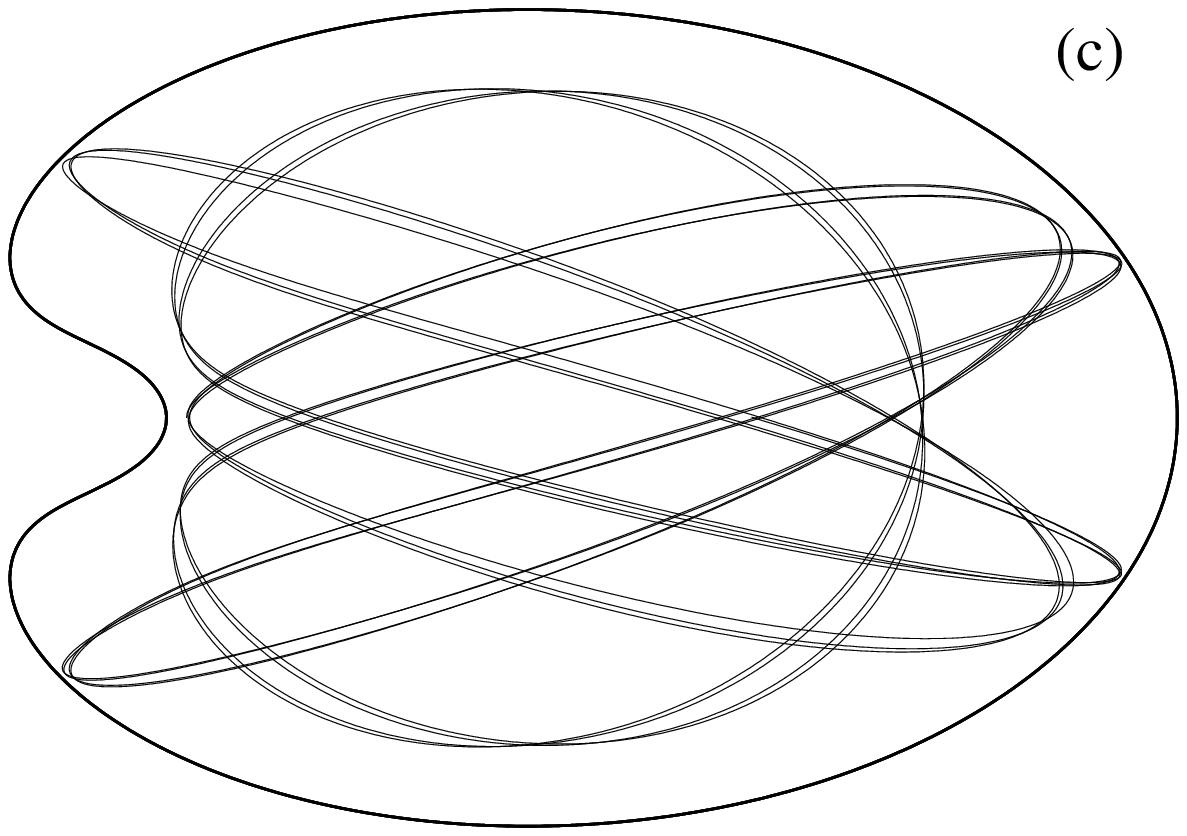} \hfill
\epsfxsize=7cm
\epsfbox{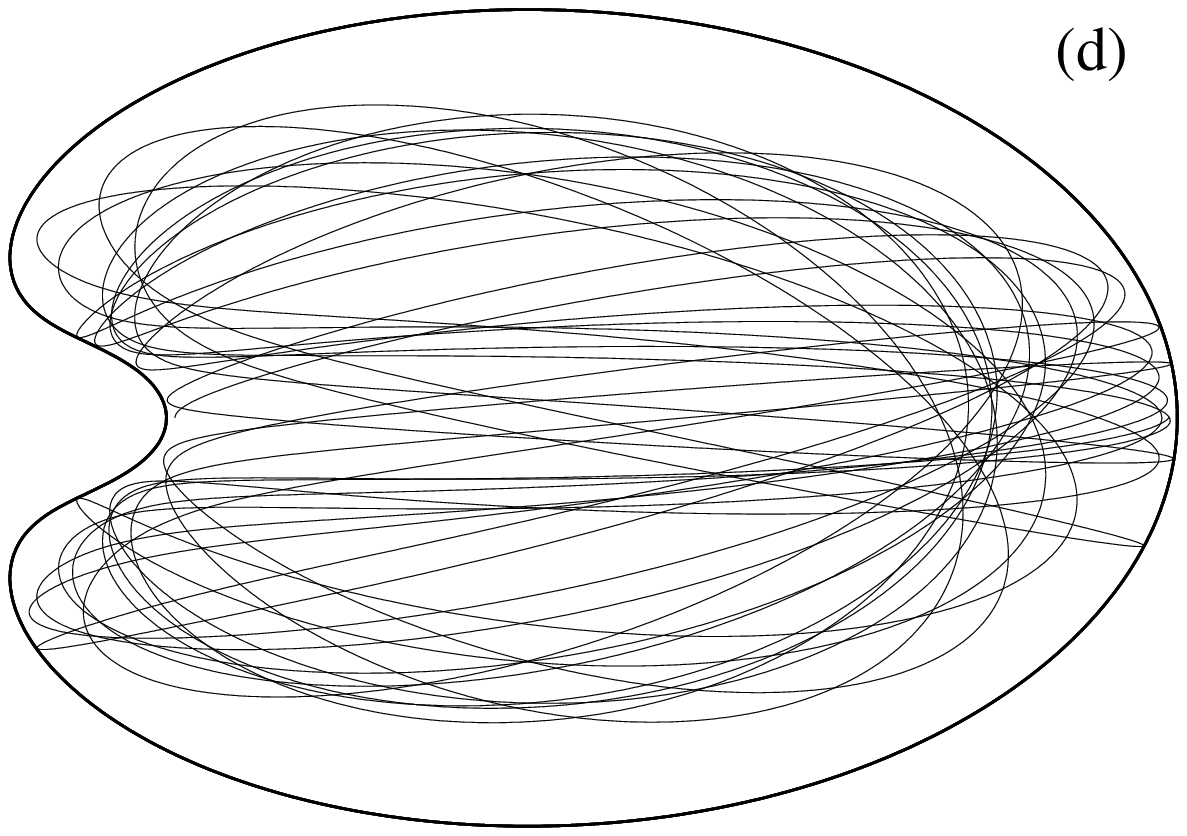}
\caption{Some trajectories in the configuration space
corresponding to the Poincare's section
of Fig. (2): (a), (b) and (c) the regular ones, corresponding to trajectories
that do not probe the 
non convex part of the boundary; (d) a chaotic one, corresponding to
a trajectory
that bounce on the non convex part of the boundary.}
\end{figure}
The behavior of the chaotic trajectories bouncing on the non convex
part of the boundary resemble classical billiards\cite{robnik}.
Two close trajectories reaching a convex part of the boundary are
bounced in a ``focusing'' way, while the non convex part defocuses
nearby bouncing trajectories, causing the divergence of the associate
geodesics, in spite of the positive Gaussian curvature $\hat{K}$.

\begin{figure}[pht]
\epsfxsize=10cm
\epsfbox{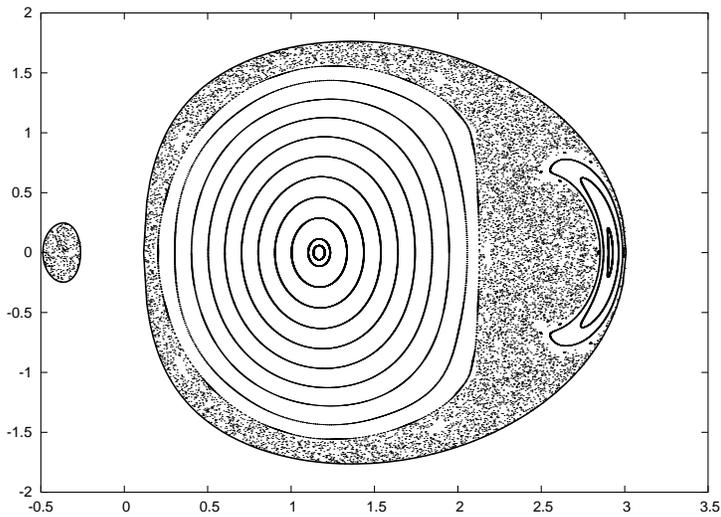}
\vspace{0.5cm}
\caption{Typical Poincar\`e's section  $(x,p_x)$ across the
plane $y=0$ for the multiply-connected ${\mathcal D}_E$ case.
For this section, one has $\alpha=1/2$ and $E/E_{\rm C}=1.02$.}
\end{figure}
Similar results hold for the multiply-connected 
${\mathcal D}_E$ case, see Fig. (4). The chaotic motion corresponds
to the trajectories that reach the non convex part of the boundary,
including its internal part (Fig. (5)). The trajectories that are
restricted to convex parts of the boundary are regular.
\begin{figure}
\epsfxsize=7cm
\epsfbox{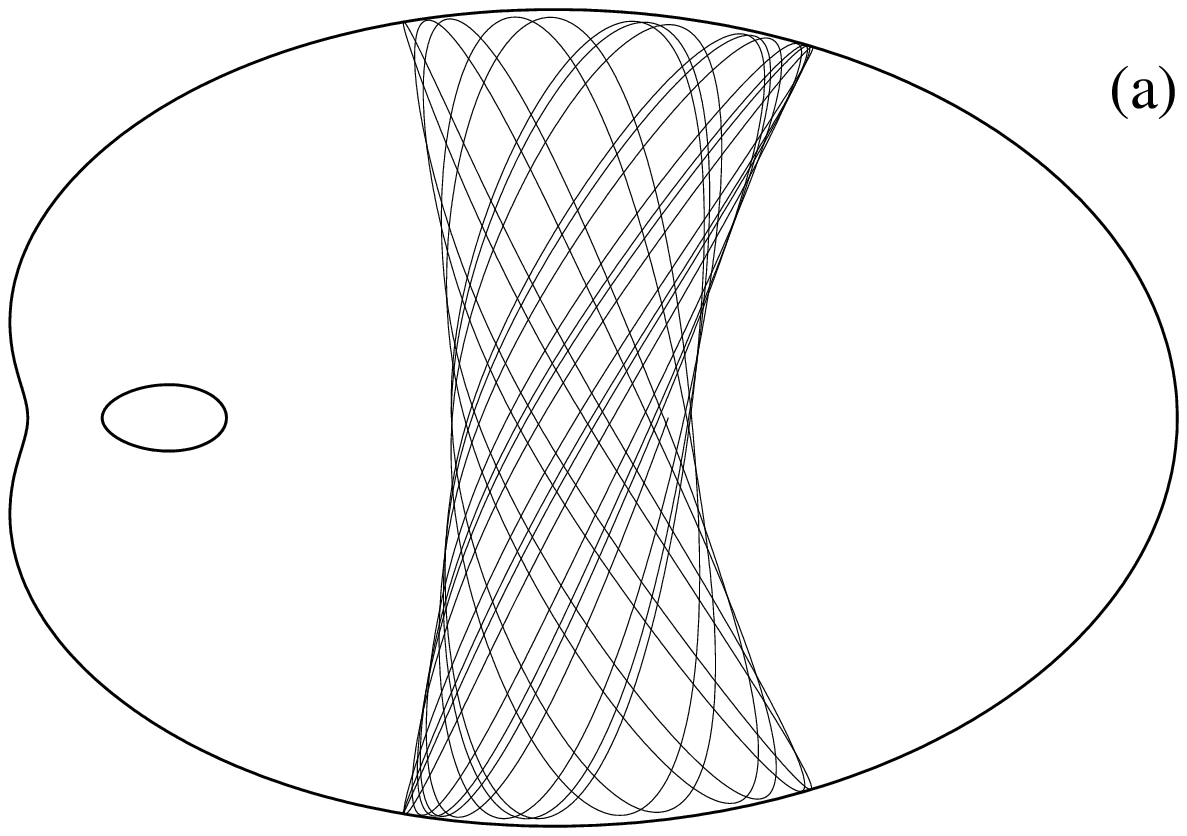} \hfill
\epsfxsize=7cm
\epsfbox{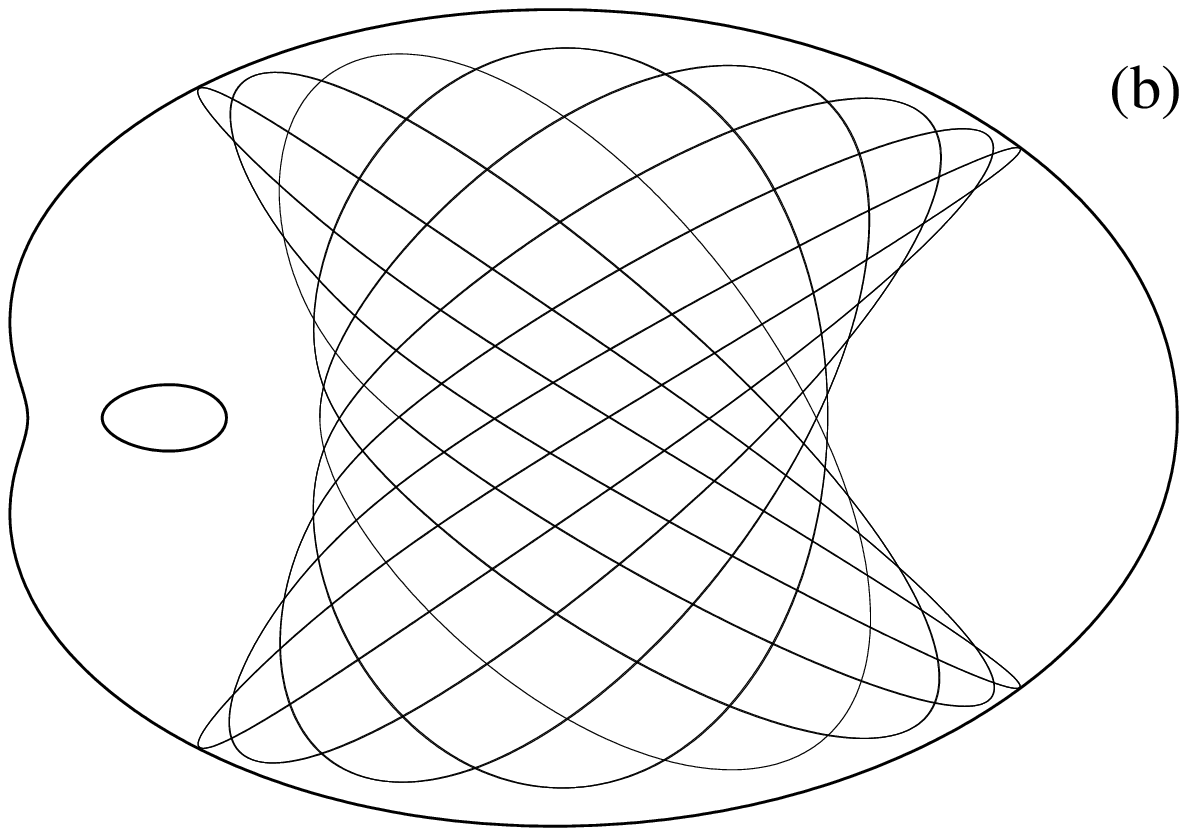} \\
\epsfxsize=7cm
\epsfbox{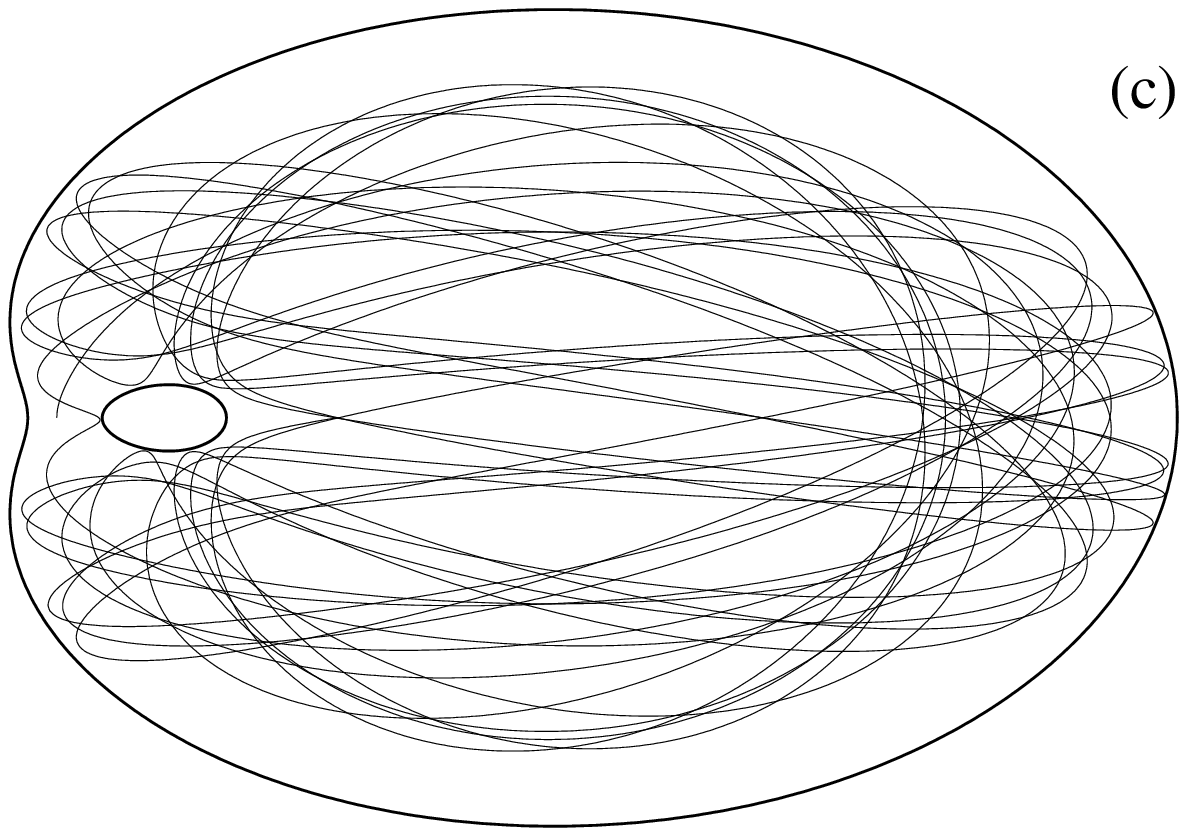} \hfill
\epsfxsize=7cm
\epsfbox{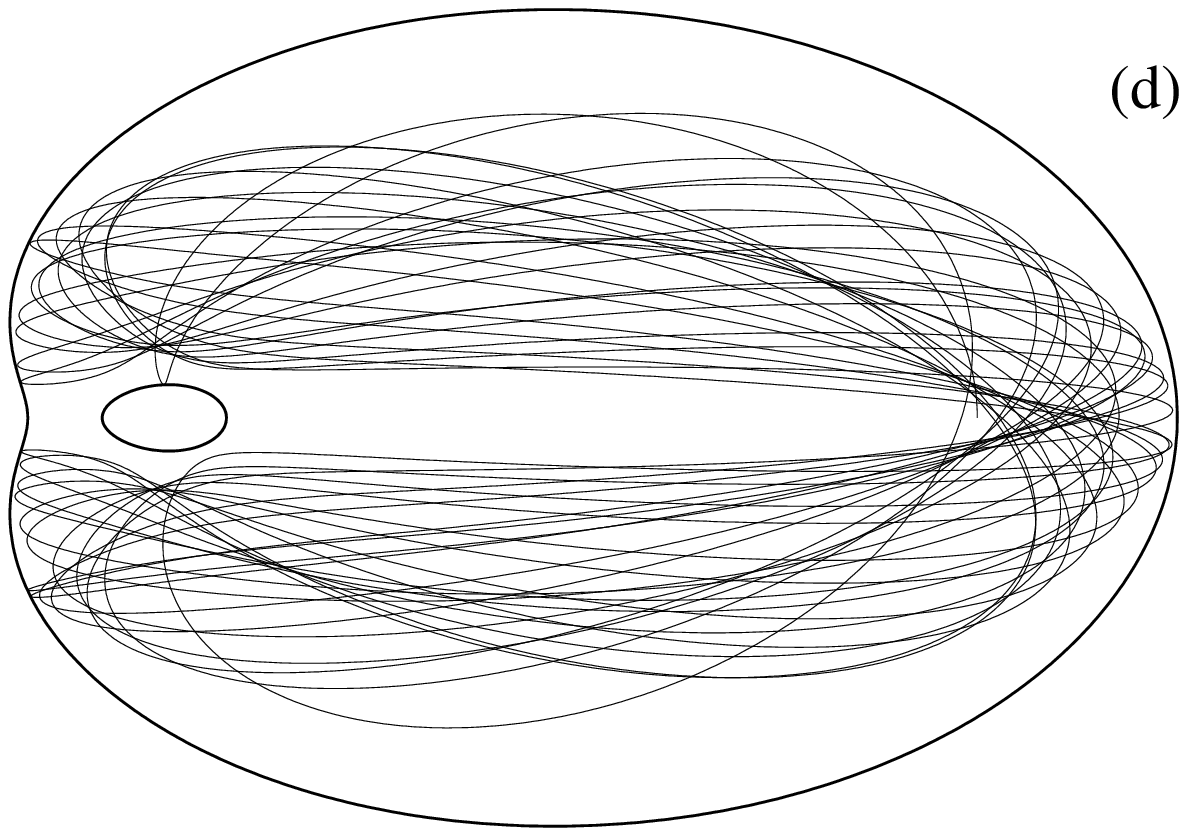}
\caption{Some trajectories in the configuration space
corresponding to the Poincare's section
of Fig. (4): (a) and (b)  the regular ones, corresponding to trajectories
that do not probe the non convex  parts of the boundary; 
(c) and (d) the chaotic ones, trajectories
bouncing on the non convex parts.}
\end{figure}

The dependency of our results on $\alpha$ shows no surprises. The smaller
is the parameter $\alpha$, the close to $E_{\rm C}$ we have the onset
of chaos. For a given $\alpha$, the value of $E/E_{\rm C}$ marking
the onset of chaos can be predicted analytically. Since the
chaotic motions arise from the bouncing on the non convex part of
the boundary, the onset of chaos will be determined by the appearance
of the non convex region, which, as one can easily see, is related
to the intersections of the curves $V(x,y)=E$ and $x^2+ y^2 = \alpha$.
These curves have two intersections in the non convex case, and no
intersections in the convex one. The value of $E$ corresponding to
only one intersection marks the transition.

The situation for the case of two or more monopoles is similar
to the expounded here. Chaotic behavior always 
appears for the trajectories that probe the non convex part of 
$\partial{\mathcal D}_{E}$, including the internal 
parts of the boundary for the multiply-connected case.

\section{Conclusion}

We have shown that the local stability criterion based on the geodesic
deviation equation for the Jacobi metric is neither necessary
nor sufficient to the occurrence of chaos. For this purpose,
we introduce
a class of 2-dimensional systems exhibiting chaotic
behavior, confirmed by numerical
evaluation of Poincar\'e sections, with everywhere positive 
Gaussian curvature, and
thus locally stable. The chaotic behavior
arises for certain trajectories bouncing on
non convex parts of 
the boundary of the effective Riemannian manifold,
resembling the behavior of some classical billiards\cite{robnik}.
Despite the fact that the positivity of the Gaussian curvature
$\hat{K}$ ensures the convergence of nearby geodesics, chaotic behavior
can arise due to defocusing of geodesics bouncing on non convex
parts of the boundary $\partial{\mathcal D}_E$.
Our result questions, once more, the viability of
local, curvature-based criteria to predict chaotic behavior.

\begin{flushleft}
\bf Acknowledgments
\end{flushleft}
The author wishes to thank H.B. Fraga for valuable discussions, the hospitality
of Prof. Enric Verdaguer from the 
Department of Fondamental Physics of the University of Barcelona, Spain,
where this work was initiated, and FAPESP for the financial support.

\end{document}